\documentclass{article}

\usepackage{amssymb,amsfonts,amsmath}
\usepackage{cite,enumerate,float}
\usepackage{color}
\usepackage{tikz}
\usetikzlibrary{arrows,snakes,backgrounds}

\def\be{\begin{eqnarray}}
\def\ee{\end{eqnarray}}
\def\nn{\nonumber}

\def\p{\partial}

\def\nKL{\widetilde{\cal KL}}
\def\KL{{\cal KL}}
\def\CK{{\cal K}}
\def\Al{{\rm Al}}  \def\Al{\Delta}

\def\ft{\mathfrak{t}}
\def\Ap{\mathfrak{A}}  \def\Ap{{\cal A}}
\def\Cp{\mathfrak{C}}  \def\Cp{{\cal C}}

\definecolor{red}{rgb}{1,0,0}
\definecolor{orange}{rgb}{1,0.5,0}
\definecolor{violet}{rgb}{0.7,0,1}

\usepackage{tikz}
\usepackage{tikz-cd}
\usetikzlibrary{arrows}
\usetikzlibrary{arrows.meta}
\usetikzlibrary{positioning}
\usetikzlibrary{shapes}
\usetikzlibrary{fit}
\usetikzlibrary{decorations.pathmorphing,decorations.pathreplacing,decorations.markings}
\usetikzlibrary{calc}

\hoffset= - 1.0in         

\begin{document}

\hfill MIPT/TH-17/26

\hfill IITP/TH-19/26

\hfill ITEP/TH-21/26

\bigskip

\bigskip

\centerline{\Large{
More on Kashaev limits of the quantum $A$-polynomials
}}

\bigskip

\bigskip

\centerline{\bf A.Morozov}

\bigskip

\centerline{ {\it MIPT, Dolgoprudny, 141701, Russia}}
\centerline{ {\it NRC ``Kurchatov Institute", 123182, Moscow, Russia}}
\centerline{ {\it Institute for Information Transmission Problems, Moscow 127994, Russia}}
\centerline{ {\it ITEP, 117259, Moscow, Russia}}

\bigskip

\bigskip

\centerline{ABSTRACT}

\bigskip

{\footnotesize
"Colored" knot polynomials satisfy difference equation w.r.t. the
highest weights of the underlying representation --
which in the case of symmetrically colored Jones are named "quantum $A$-polynomials".
In the double scaling quasiclassical (Kashaev) limit, when representation size $r\sim \hbar^{-1}$,
there are different phases -- in one of them the classical action vanishes
and in another one it is a deformation of hyperbolic volume (of a knot complement in $S^3$).
This corresponds to a splitting of the non-homogeneous version of the quantum $A$-polynomial
into two pieces, which we illustrate by more examples than just a figure-eight knot $4_1$
in the original paper \cite{GMkl}.
From the point of view of quasiclassics, hyperbolic volume is just an integration constant,
which is not fully determined by the $A$-polynomial equation -- and actually remains ambiguous
in this formalism.
As a byproduct, we expect that classical $A$-polynomial at $L=1$ becomes proportional to Alexander:
$A^{\cal K}(1,M)\sim \Delta^{\cal K}(M)$ -- this seems true, but $A$
should be consistent with the polynomiality of {\it non-homogeneous quantum} $\Ap$-polynomial,
what sometime implies that it is not minimal.

}

\bigskip

\bigskip

\section{Introduction}

Knot polynomials are usually interpreted as Wilson averages in one of the two ($2d$ and $3d$)
simplest Yang-Mills theories, which do not contain particles -- only topological degrees of freedom.
In the case of knot polynomials this is $3d$ Chern-Simons (CS) theory \cite{CS}.
Here one pays a lot of attention to different representations --
which would be equally interesting in $4d$ QCD-like models
(where, on one hand, it would distinguish between confined and deconfined averages,
i.e. between the area and perimeter laws, while on another hand it would help to understand
the hidden integrability structure of non-perturbative calculus {\it a la} \cite{UFN3}),
but is not yet studied there in enough detail.
In CS it is -- and one of the possible ways to control representation dependence
is through equations with respect to representation labels.
In the case of symmetric representations the label is a single $r$,
and the corresponding difference equation is known as {\it quantum $A$-polynomial} 
\cite{qAp,MMeqknpols,GMqap},
because its classical version at $q=1$ appears to reproduce the ordinary $A$-polynomial,
which has various interpretations in terms of different topological structures \cite{Ap}.
Restriction to symmetric representations is natural for the $su_2$ gauge group,
thus the story of $A$-polynomials refers usually to Jones polynomials \cite{Jones}
rather than more general HOMFLY-PT \cite{HOMFLY-PT}, Kauffman \cite{Kauffman} etc.
Besides generalizations to non-symmetric representations,
there are two other directions -- to $C$-polynomials \cite{Cpols,MMcpols} instead of $A$
(they take into account the differemtial/cyclotomic expansion (DE) \cite{DE} -- another peculiarity
of $r$-dependence, what allows to somehow reduce the role of field-theory structures and
make equations more focused on the properties of particular knot)
and to superpolynomials (Khovanov and Khovanov-Rozansky homologies \cite{KhR}) instead of
Jones and HOMFLY-PT/Kauffman.

Still another option, when we have a parameter, like $r$, is to study the limits,
say $r\longrightarrow \infty$.
It is nicely-defined in the double-scaling version, when also $q\longrightarrow 1$,
when the surviving parameter is $\ft:=q^r$.
We call it Kashaev limit ($\KL$), referring to the breakthrough discoveries, made here by R.Kashaev
in mid-90's \cite{Kashaev}.
In a recent \cite{GMkl} we suggested that the quantum $A$-polynomial for $r$-colored Jones
 acquires the peculiar form
\begin{equation}\label{main2}
\boxed{
	\begin{array}{c}
		\begin{tikzpicture}
			\node {$\left(\Ap^{\CK}\left(e^{S'(\ft)},\ft\right)-\Ap^{\CK}\left(1,\ft\right)\right)J^{\CK}+
Q^\CK(\ft)\left(\Al^{\CK}(\ft)J^{\CK}-1\right)=O(\hbar)\,.$};
			\foreach \x/\y/\h in {-4.8/2.6/0.3} {
				\draw[ultra thin, fill] (\x,\h) to[out=90,in=180] (\x+0.3,\h+0.17) -- (0.5*\x+0.5*\y-0.3,\h+0.17) to[out=0,in=270] (0.5*\x+0.5*\y,\h+0.3) to[out=270,in=0] (0.5*\x+0.5*\y-0.3,\h+0.13) -- (\x+0.3,\h+0.13) to[out=180,in=90] (\x,\h);
				\draw[ultra thin, fill] (\y,\h) to[out=90,in=0] (\y-0.3,\h+0.17) -- (0.5*\x+0.5*\y+0.3,\h+0.17) to[out=180,in=270] (0.5*\x+0.5*\y,\h+0.3) to[out=270,in=180] (0.5*\x+0.5*\y+0.3,\h+0.13) -- (\y-0.3,\h+0.13) to[out=0,in=90] (\y,\h);
				\node[above] at (0.5*\x+0.5*\y,\h+0.3) {$\scriptstyle 
\Ap^{\CK}\left(e^{S'(\ft)},\ft\right)-\Ap^{\CK}\left(1,\ft\right)+Q^\CK(\ft)\Al^{\CK}(\ft)
\ =\ 0$};
			}
			\foreach \x/\y/\h in {1.1/3.4/-0.2} {
				\draw[ultra thin, fill] (\x,\h) to[out=270,in=180] (\x+0.3,\h-0.17) -- (0.5*\x+0.5*\y-0.3,\h-0.17) to[out=0,in=90] (0.5*\x+0.5*\y,\h-0.3) to[out=90,in=0] (0.5*\x+0.5*\y-0.3,\h-0.13) -- (\x+0.3,\h-0.13) to[out=180,in=270] (\x,\h);
				\draw[ultra thin, fill] (\y,\h) to[out=270,in=0] (\y-0.3,\h-0.17) -- (0.5*\x+0.5*\y+0.3,\h-0.17) to[out=180,in=90] (0.5*\x+0.5*\y,\h-0.3) to[out=90,in=180] (0.5*\x+0.5*\y+0.3,\h-0.13) -- (\y-0.3,\h-0.13) to[out=0,in=270] (\y,\h);
				\node[below] at (0.5*\x+0.5*\y,\h-0.3) {$\scriptstyle \Al_{\CK}(\ft)J_{\CK}=1 $};
			}
		\end{tikzpicture}
	\end{array}
}
\end{equation}
where $\Al^{\cal K}(\ft):=H^{\cal K}_\Box(A=1,q=\ft)$ is the Alexander polynomial, i.e. HOMFLY at $N=0$.

Actually there are two phases, i.e. two different Kashaev limits:
the underbraced  one, which we call ``naive'', where
\be
\nKL(J^\CK) = \Al^{\CK}(\ft)^{-1}
\label{inverAl}
\ee
and the overbraced ``true'' one, where $\KL(J^\CK)$ behaves quasiclassically
\be
\KL(J^{\CK}) \sim e^{S(\ft)/\hbar}
\ee
and
\be
S(\ft=1) = {\rm hyperbolic\ volume\ of}\ S^3/\CK
\label{hypvol}
\ee
Eq.(\ref{inverAl}) is the celebrated observation of P.Melvin and H.Morton   \cite{MM,KLM},
usually proved/explained with the help of Burau formalism \cite{Burau},
while eq.(\ref{hypvol}) is the equally celebrated result of R.Kashaev \cite{Kashaev} --
and (\ref{main2})  unifies them in a simple and transparent way.
The only thing, which is lacking, is the clear understanding of (\ref{main2}) itself
or at least enough evidence that it really holds in this form for considerable set
of knots.
In \cite{GMkl} we actually studied a single example of $\CK=4_1$,
and now we slightly  extend it to  some other knots.

\bigskip

There are several aspects,
which deserve better understanding:

\begin{itemize}

\item{}
Eq.(\ref{main2}) unifies two different Kashaev limits,
but in fact it does not reveal the most interesting feature (\ref{hypvol}).
The $A$-polynomial equation (\ref{main2}) explains how the classical action depends on $\ft$,
but it is not sensitive to the $\ft$-constant term (\ref{hypvol}).
Additional principle for its extraction from (\ref{hypvol}) requires the ``lifting rule'',
i.e. the requirement that solutions have peculiar ``hypergeometric'' form/shape.
What adds to the mystery, is that the equation may be not fully sensitive to embedding into exactly $S_3$
-- its derivation in \cite{GMqap} involves global transformation of contours, but it is not yet clear
if exactly $S_3$ is needed to allow them, and which.
Thus this particular embedding can not fully fix integration constant, associated with this particular case.

\item{}
Usually the derivation of (\ref{hypvol}) requires detailed knowledge about the $q$-dependence
of the DE coefficients $a_s(q)$, to be introduced in sec.\ref{nKL} below,
while in (\ref{main2}) distinguished is the role of Alexander, which depends only
on their values $a_s(1)$ at a particular point $q=1$.
Exact difference between information, contained in sophisticated $a_s(q)$ and simple $a_s(1)$
remains obscure.

\item{}
For $4_1$ the $C$-polynomials are trivial, so this case
a kind of defines the transformation/rotation  between $\Cp$- and $\Ap$-polynomials,
which can be further used to transform the basic (and simpler) $\Cp$-pols
into less trivial $\Ap$-pols.
Of particular interest is Kashaev limit of this transformation.

\item{}
Kashaev limits can be described by peculiar $R$-matrices.
In the case of $\nKL$ they were discussed in \cite{KLM}, it is naturally expected that for Jones
such $R$-matrix would be related to Burau one, which nicely describes inverse Alexander,
while its generalizations to $N>2$ is a partly open problem.
In the case of $\KL$ the hyperbolic volumes are known to be nicely described \cite{Kashaev} by the
quantum-dilogarithm $R$-matrix \cite{FK}.
It is a question what is the proper interpolation between Burau and dilogarithm $R$-matrices,
which could stand behind the l.h.s of the equation (\ref{main2}) -- even for Jones.
More interesting/intriguing  is interpretation of all these three limiting $R$-matrices
in terms of the   universal one \cite{univR} -- what are the representations,
responsible for them.

\end{itemize}

We do not provide exhaustive comments on these additional issues in the present text,
still it can help to highlight the intrigue.

\bigskip

The structure of the paper is as follows.

We begin in sec.\ref{DE} from a short reminder of differential expansion \cite{DE}.
Formally it is not needed for  neither Kashaev limit nor $A$-polynomials.
However, it is convenient for both and most existing considerations use it, explicitly or implicitly.
In the case of Jones it substitute colored Jones by a sequence of functions $a_s(q)$.

The Melvin-Morton conjecture, surveyed in the next sec.\ref{nKL},
expresses their values at $q=1$  through Alexander polynomial $\Delta(\ft)$,
which is a polynomial of finite degree, i.e. contains just a finite number of coefficients,
which depends only on the defect.
Thus for a given defect only finite number of $a_s(1)$ is independent,
i.e. there is a kind of a peculiar sum rule for the set $\{a_s(1)\}$.

In sec.\ref{tKL} we switch to the ''true'' Kashaev limit \cite{Kashaev},
which involves an accurate quasiclassical limit $q=e^\hbar$, $\hbar\longrightarrow 0$,
taking into account the true behavior of the coefficients $a_s(q)$.
In the double scaling limit, when $\ft =q^r$ is kept finite, the DE series
behaves as $\KL(J^\CK)\sim\frac{\exp(S_{\cal K}/\hbar)}{\sqrt{\det S_\CK''}}$,
where $S_\CK$ can be non-vanishing -- unlike it was assumed to be in the ''naive'' limit.
As usual in quasiclassics it is defined ambiguously, but there is always a possibility (a branch choice)
to make it matching the hyperbolic volume of the complement $S^3/\CK$ --
e.g. to define the limit/branch as coinciding with Hikami calculus \cite{Hikami}.
In sec.\ref{tKL} we illustrate this with the simplest example of the figure-eight knot $\CK=4_1$,
where the branch is actually unique and the limit is defined in the most natural way.

In sec.\ref{Apols} we come closer to the main subject of this paper.
We consider the simplest examples of quantum $\Ap$-polynomials -- for the trefoil $\CK= 3_1$
and for the figure-eight knot $\CK=4_1$.
These are finite-difference equations in $r$, and they have a peculiar decomposition,
with the two components relevant for description of quasiclassics with vanishing and non-vanishing
classical action -- which correspond to the cases of ``naive'' and ``true'' Kashaev limits.
One more  limit provides a classical $A$-polynomial.
The subsection \ref{mainclaim} formulates our main claims: decomposition (\ref{main2}) \cite{GMkl}
of the non-homogeneous quantum $\Ap$-polynomial and divisibility of classical $A$-polynomial
by Alexander (with certain reservations).

The next two sections \ref{exaAp} and \ref{clAp} contain simple illustrations to these
main claims with the help of various small knots.

A very short sec.\ref{Cpols} is a brief comment on the possible role of quantum $\Cp$-polynomials.
They are equations on the coefficients $a_s(q)$ instead of $\Ap$-polynomials which are imposed on $J_r$.
In particular they get trivial for figure-eight, since $a_s^{4_1} = 1$, so that figure-eight
a kind of intertwines $\Cp$ and $\Ap$ polynomials or transforms ones into the others.
As already mentioned, exact description of this transformation is beyond the scope of the present paper
and is left for the future.

\section{Differential expansion  \label{DE}}

DE \cite{DE} of symmetric HOMFLY for generic knot  $\CK$ separates the dependencies on the knot ${\cal K}$
and representation $[r]$:
\be
H^{\cal K}_{[r]}(q,A) = 1 + \{A/q\} \sum_{s=1}^n  G_s^{\cal K}(A,q)\cdot \frac{[r]!}{[s]![r-s]!}
\prod_{j=0}^{s-1} \{Aq^{r+j}\}
\ee
In the framework of the Reshetikhin-Turaev formalism \cite{RT} it is basically a consequence of
elementary group theory \cite{BiMo}.
The coefficients $G_s^{\cal K}$ are often factorized further, but the depth of factorization
depends on the knot or, more specifically, on its single integer-valued characteristic,
named {\it defect} $\delta^\CK$  in \cite{Konodef}.
Defect also defines the power  of Alexander, $\Al^{\cal K}(\ft) = H_\Box^{\cal K}(A=1,q=\ft)$:
it is $1+\delta^\CK$, if Alexander is considered as a function of symmetric variable $\{\ft\}^2$.

As explained in \cite{KLM}, DE implies for (normalized) Jones that
\be
J_r^\CK(q) := H^\CK_{[r]}(q,A=q^2)
= 1 + \sum_{s=1}^r a_s^\CK(q)\prod_{\stackrel{\j\neq 0}{j=-s}}^s \{q^{r+j+1}\}
\label{DEJones}
\ee
Defect  affects the definition of $a_s(q)$ in original terms of knot theory,
\be
a_s^{\cal K}(q)= \frac{G_s^{\cal K}(q^2,q)}{[s]!\{q\}^{s-1}}
\label{avsG}
\ee
however it no longer matters if we consider $a_s(q)$ as a given entity -- as we do in this paper.

\section{The naive Kashaev limit and a miraculous sum rule
\label{nKL}}

What we call "naive" Kashaev limit, is the formal substitution of $q^r=\mathfrak{t}$
and $q=1$ into the coefficients of the DE (\ref{DEJones}) --
{\it disregarding} the possibility of accumulation of many $r$-independent terms,
what could produce additional $\ft$-dependence,
like in $\sum_{s=0}^r q^s = \frac{1-q\ft}{1-q}$.
This is an elementary operation:
we substitute $\{q^{r+j+1}\}\longrightarrow \{\ft\}$ and $a_s(q)\longrightarrow a_s(1)$,
into  (\ref{DEJones}) and obtain
\be
\nKL(J^\CK) =1+ \sum_{s=1}^\infty a^\CK_s(1) \{\mathfrak{t}\}^{2s}
\label{nKLJ}
\ee
Once DE is explicitly known (what is true for many types of knots),
one can check that this series always satisfies
\be
\nKL^\CK(\ft) = \frac{1}{\Al^{\CK}}
\label{MMconj}
\ee
This is the Melvin-Morton conjecture, studied and proved in \cite{MM}.
It has a direct generalization to HOMFLY and Kauffman polynomials, see \cite{KLM}.

For Alexander polynomial the implication of DE is rather trivial:
\be
\Al^\CK(\ft) := H^\CK_{[1]}(A=1,q=\ft) = 1 -  f^\CK_1(\ft) \{\ft\}^2
\label{DEAl}
\ee
Still, according to (\ref{MMconj}), the two expressions (\ref{nKLJ}) and (\ref{DEAl})
are inverse to each other, i.e. all $a^\CK_s(1)$ are made from $f^\CK_1(\ft)$ and vice versa.
The coefficient $F_1(\ft)$ is actually symmetric under the change $\ft\leftrightarrow -\ft^{-1}$
(such changes correspond to the transposition of Young diagram, which leaves $[1]=\Box$ intact)
and thus actually depends on $\{\ft\}^2$.
Both $\{a^\CK_s(1)\}$ and $\{f^\CK_1(\ft)\}$
are one-parametric sets (labeled respectively by $s$ and by powers of $\{\ft\}^2$),
thus the amount of information matches.
Still it is a non-trivial sum rule for $a_s(1)$, especially because the power of Alexander
(as a function of $\{\ft\}^2$) is finite -- equal to $\delta^\CK +1$ \cite{Konodef}.

Note that in terms of original DE
\be
H_{[r]}^\CK(q,A) = 1 + [r] F^\CK_1(A,q)\{A/q\}\{Aq^r\} + \ldots
\ee
where the higher terms depend on the defect, the first coefficient
$a^\CK_1(q)=F^\CK_1(q^2,q)$ while $f_1^\CK(\ft) = F^\CK_1(1,\ft)$, so they look like two independent functions.
However, due to (\ref{MMconj}), the values $a^\CK_s(1)$ at $q=1$ and $f_1^\CK(\ft)$ are   not independent
and are related in a knot-independent way.
A trivial particular case is $a_1^\CK(1) = f^\CK_1(1)$ -- this is obvious because both are equal to $F_1^\CK(1,1)$.
However, relations between $a_s(1)$ with $s>1$ do not look so evident.

\section{The true Kashaev limit of Jones polynomials
\label{tKL}}

\subsection{Outline}

The true Kashaev limit corresponds to evaluation of Chern-Simons functional integral
over the complement of the knot $S^3/{\CK}$,
its localization on the specific hyperbolic connection ${\cal A}_{\rm hyp}^{\cal K}$
and evaluation of the result in the form
\be
\KL({\rm Jones}^{\cal K}) = \frac{e^{\frac{\rm action}{\hbar}}}{\sqrt{\rm Det}}
\longrightarrow   \exp\left(\frac{{\rm hypVolume}}{\hbar}+O(1)\right)
\ee
In fact $\KL({\rm Jones}^{\cal K})$ is still a function of $\ft$.
Hyperbolic volume appears from it in the special point $\mathfrak{t}=1$,
when the connection has no monodromies around the knot,
i.e. the knot is smoothly embedded.
From the point of view of the equation w.r.t. $r$ or $\ft$ the hyperbolic volume is
just an integration constant, which is not fixed unambiguously by this equation.
It can be extracted by the study of original series (differential expansion \cite{DE})
or by imposing some equivalent restrictions on the solution.
Still, as we shall see, the $\ft$-dependence of the action $S(\ft)$ is
quite non-trivial -- and can be considered as a corollary of the
overbraced piece of the equation (\ref{main2}).

\subsection{Figure eight knot $4_1$
\label{41main}}

\subsubsection{DE}
Differential/cyclotomic expansion (DE) \cite{DE} for $4_1$ is especially simple:
all the coefficients $a^{4_1}_s = 1$:
\be
J_R^{4_1} = \frac{1}{\{q^{r+1}\}} \sum_{s=0}^\infty \prod_{j=-s}^s \{q^{r+j+1}\}
\label{DE41}
\ee
The sum over $s$ is actually cut-off at $s=r$, because   the products are vanishing for $s>r$.

\subsubsection{Quasiclassical limit: from $q$ to $\bf t$ and from sum to an integral
\label{QcL41}}
It can be rewritten in a form close to the usual quasiclassics in at least two complimentary ways:
\be
\sum_{s=0}^\infty \prod_{j=-s}^{s} \{q^{r+j+1}\} =
\sum_{s=0}^\infty  q^{(r+1)(2s+1)}\prod_{j=-s}^s (1-q^{-2r-2j-2}) =
\sum_{s=0}^\infty q^{(r+1)(2s+1)}
\frac{(q^{-2r-2s-2}|q^2)_\infty}{(q^{-2r+2s}|q^2)_\infty}
\label{S41oraa}
\ee
where the summand in Kashaev limit with finite $\ft:=q^{r}=e^{\hbar r} $ and $z:= q^{2s}=e^{\hbar s} $
and $q^{2rs} = e^{\frac{1}{\hbar}\log \ft\cdot\log z}$  becomes
\be
q^{(r+1)(2s+1)}
\frac{(q^{-2r-2s-2}|q^2)_\infty}{(q^{-2r+2s}|q^2)_\infty}
\longrightarrow \exp\left(\frac{1}{2\hbar}
\Big({\rm Li}_2\left(\frac{1}{\ft^2 z }\right) - {\rm Li}_2\left(\frac{z}{\ft^2}\right)+2\log\ft \cdot\log z \Big)
\right)
=\nn \\
=  \exp\left(\frac{1}{2\hbar}
\left\{L\left( \frac{1}{\ft^2 z }\right) - L\left(\frac{z}{\ft^2}\right)+2\log\ft \cdot\log z
-\frac{1}{2}\left(\log \frac{1}{\ft^2 z }\cdot \log\left(1-\frac{1}{\ft^2 z }\right)
- \log(\frac{z}{\ft^2})\cdot \log\left(1-\frac{z}{\ft^2}\right)\right)\right\}\right)
\label{S41aa}
\ee
and/or
\be
\sum_{s=0}^\infty \prod_{j=-s}^{s} \{q^{r+j+1}\} =
\sum_{s=0}^\infty  q^{-(r+1)(2s+1)}\prod_{j=-s}^s (q^{2r+2j+2}-1) =
\sum_{s=0}^\infty q^{-(r+1)(2s+1)}\frac{(q^{2r-2s+2}|q^2)_\infty}{(q^{2r+2s+4}|q^2)_\infty}
\label{S41orbb}
\ee
\be
q^{-(r+1)(2s+1)}\frac{(q^{2r-2s+2}|q^2)_\infty}{(q^{2r+2s+4}|q^2)_\infty}
\longrightarrow \exp\left(\frac{1}{2\hbar}
\left({\rm Li}_2\left(\frac{\ft^2}{z}\right) - {\rm Li}_2(\ft^2 z)-2\log\ft \cdot\log z\right)\right)
=\nn \\
=  \exp\left(\frac{1}{2\hbar}
\left\{L\left(\frac{\ft^2}{z}\right) - L(\ft^2 z)-2\log\ft \cdot\log z
-\frac{1}{2}\left(\log \frac{\ft^2}{z}\cdot \log\left(1-\frac{\ft^2}{z}\right)
- \log(\ft^2 z)\cdot \log(1-\ft^2 z)\right)\right\}\right)
\label{S41bb}
\ee
Here we used the quasiclassical approximation to Pochhammer symbol
\be
(x|q^2)_\infty := \prod_{k=0}^\infty (1-xq^{2k})
= \exp\left(-\sum_{n=1}^\infty \frac{x^n}{n(1-q^{2n})}\right)
= \sqrt{1-x}\cdot\exp\left(\frac{1}{2\hbar}{\rm Li}_2(x) + O(1)\right)
\ee
where
\be
{\rm Li}_2(x):=\sum_{n=1}^\infty \frac{x^n}{n^2}
\ \ \ \Longrightarrow \ \ \
\p_x {\rm Li}_2(x) = -\frac{\log(1-x)}{x}
\label{LiL}
\ee
and
\be
L(x):={\rm Li}_2(x) + \frac{1}{2}\log x \cdot \log(1-x)
\ee
The two expressions (\ref{S41aa}) and (\ref{S41bb})
should be understood in the sense of analytical continuations, and they
coincide due to the simplest dilogarithm identity
\be
L(x) + L(x^{-1})=\frac{\pi^2}{3}
\label{L+Linv}
\ee
Note that while (\ref{S41aa}) and (\ref{S41bb}) coincide, there ancestors need more care:
in (\ref{S41oraa}) there stands a ratio of two products which have vanishing factors for $s=r-1$
in the numerator and $s=r$ in denominator.
The two zeroes annihilate each other, still attention is needed.
To the contrary, (\ref{S41orbb}) is free of this ``problem'' --
which fully disappears in (\ref{S41aa})=(\ref{S41bb}).

\bigskip

Changing the sum over $s$ for
integral over $dz/z$,
we obtain in the quasiclassical limit of $\hbar\longrightarrow 0$,
when the pre-exponentials can be also neglected:
\be
J_r^{4_1} = \int \frac{\exp\left(\frac{S(z,t)}{\hbar}\right)}{\sqrt{\p^2_z S(z,t) }}\frac{dz}{z}
\cdot\Big(1+  O(\hbar)\Big)
\ee
with
\be
S(z,\mathfrak{t}) = \log \,\mathfrak{t}  \cdot \log z
+ \frac{1}{2}\left({\rm Li}_2\left(\frac{1}{z\mathfrak{t}^2}\right)
- {\rm Li}_2\left(\frac{z}{\mathfrak{t}^2}\right)\right)
\label{S41}
\ee

\subsubsection{Saddle point}

Saddle point $z_*(\ft)$ is at $\p_z S = 0$, i.e.   at
\be
(\mathfrak{t}^2-z_*)(\mathfrak{t}^2z_*-1) = \mathfrak{t}^2z_*
\label{z*41}
\ee
and we get
\be
J_R^{4_1} =
\exp\left(\frac{S(z_*,t)}{\hbar}\right) +  O(1)
\ee

One can also represent it as $S(z_*,t) = \int \frac{d\mathfrak{t}}{\mathfrak{t}}\,\log \,g(\mathfrak{t})$
where $g(\mathfrak{t}) = e^{\mathfrak{t}\,d_\mathfrak{t} S(z^*,\mathfrak{t})}$.
From (\ref{S41}) we obtain:
\be
\mathfrak{t}\,d_\mathfrak{t} S(z^*,\mathfrak{t})
= \mathfrak{t}\,\p_\mathfrak{t} z_* \underbrace{\p_z S(z^*,\mathfrak{t})}_{0\ {\rm  at}\ z_*}
+ \mathfrak{t}\,\p_\mathfrak{t} S(z^*,\mathfrak{t})
=  \mathfrak{t}\cdot\Big(\frac{\log z_*}{\mathfrak{t}}+
\frac{1}{z_*\mathfrak{t}^3}
\underbrace{z_*\mathfrak{t}^2 \log(1-1/z_*\mathfrak{t}^2)}_{{\rm Li}_2^\prime\left(\frac{1}{z\mathfrak{t}^2}\right)}
-\frac{z_*}{\mathfrak{t}^3}
\underbrace{\frac{\mathfrak{t}^2}{z_*} \log(1-z_*/\mathfrak{t}^2)\Big)
}_{ {\rm Li}_2^\prime\left(\frac{z}{\mathfrak{t}^2}\right)} = \nn
\ee
\be
  = \log \frac{1-\mathfrak{t}^2z_*}{z_*-\mathfrak{t}^2}
\ee
so that $g(\mathfrak{t}) = \frac{1-\mathfrak{t}^2z_*}{z_*-\mathfrak{t}^2}$ or
 $z_*=\frac{g\mathfrak{t}^2+1}{g+\mathfrak{t}^2}$.
Substitution into (\ref{z*41}) implies a quadratic equation for $g$:
\be
g^2 + (2+\mathfrak{t}^2+\mathfrak{t}^{-2} -\mathfrak{t}^4-\mathfrak{t}^{-4})\cdot g +1 = 0
\label{geq41a}
\ee

\subsubsection{Hyperbolic volume}

At $\mathfrak{t}=1$ the action (\ref{S41}) turns into
\be
S(z_*,\mathfrak{t}=1) = \frac{1}{2}\left({\rm Li}_2\left(\frac{1}{z_* }\right)- {\rm Li}_2(z_*)\right)
\ \ \ {\rm at} \ \ \
z_*-1+z_*^{-1} = 0 \ \ \ {\rm i.e.} \ \ \ z_* = e^{\pi i/3}
\ee
Then
\be
i S^{4_1}(e^{\pi i/3},\mathfrak{t}=1) = \frac{i}{2}\left({\rm Li}_2\left(\frac{1}{z_* }\right)- {\rm Li}_2(z_*)\right)
= 1.014941607\ldots
\label{hypvol41}
\ee
is a hypervolume of $S^3/3_1$.

Note that (\ref{hypvol41}) is a corollary of  (\ref{S41}) and {\it can not} be derived from  (\ref{main2}),
which is {\it another} corollary of (\ref{S41}).
The hypervolume is the $\ft$-independent part of the classical action can be extracted only
from the original DE expression
for $J_r$, and is {\it not} controlled by the $\Ap$-polynomial,
which knows much more -- but only about the $r$-{\it dependence} of $J_r$.

\subsection{All twist knots $\CK=[2m,1]$ and $\CK=[2m+1,2]$}

To provide a simplest illustration of what happens in generalizations,
we consider just the very simple extension -- to arbitrary twist knots.
The differ from $4_1$ by non-triviality of the coefficients $a_s(q)$ --
and in variance from torus knots they are hyperbolic, thus have
non-trivial saddle points and hyperbolic volume.
Instead there are different realizations of $a_s(q)$,
thus different integral representations, even with different numbers
of integration variables,
thus different actions and different saddle points.
Only few of them correspond to the hyperbolic volume of $S^3/\CK$,
and one needs additional arguments to extract the true one.
An option is to use dedicated formalisms (choice of $R$-matrices
in RT approach), like corresponding to gluing of tetrahedra --
but ambiguities often persist even there, and additional requirements like
positivity of gluing angles should be imposed.

In this paper we do not go into too many details and exploit just the
simplest single-integral description of $a_s^{tw_K}$.
Thus the only source of ambiguity will be that in the choice of saddle points,
still it is enough to illustrate the problem.

\subsubsection{Generic $a^{tw_K}$ and an example of $\Cp$-polynomial}

According to \cite{evo}
\be
a^{tw_k}_s(q) =
\left.\frac{F_s^{tw_k}(A,q)}{[s]!\{q\}^{s-1}}\prod_{i=0}^{s-2}\{Aq^{i }\}\right|_{A=q^2}
= F_s^{tw_k}( q^2,q)
=q^{\frac{s(s+3)}{2}}\sum_{j=0}^s \frac{(-)^jq^{2j(j+1)k}}{[j]![s-j]!}\frac{\{q^{2j+1}\}}{\{q\}^{s-1}}
\frac{\prod_{i=1}^{s-2} \{q^{i+2}\}}{\prod_{i=j-1}^{s+j-1} \{q^{i+2}\}}
= \nn \\
=\frac{q^{\frac{s(s+3)}{2}}}{\{q\}^s}
\sum_{j=0}^s (-)^jq^{2j(j+1)k}\frac{[s]![2j+1]}{[s-j]![s+j+1]!}
\label{atwk}
\ee
They satisfy their own equations (described by $n$-evolution or by so called $\Cp$-polynomials \cite{MMcpols}),
which for $k=-2$ (i.e. for $\CK=6_1$) look like
\be
F_{r+1}^{6_1} - \left(1+\frac{1}{A^2q^{2r}}\right)F_r^{6_1} = q^{-2}(1-q^{-2r})
\left(\left(1-\frac{1}{A^2q^{2r-2}}\right)F_r^{6_1} - F_{r-1}^{6_1}\right)
\ee
Complexity of $\Cp$-polynomials  grows with $k$.

\subsubsection{Quasiclassics: additional integration and a variety of saddle points}

Since (\ref{atwk}) contains a new sum -- over $j$ --
we get a ne integration variable
$u =q^{2j}=e^{2\hbar j}$  in addition to $z=q^{2s}=e^{2\hbar s}$:
\be
a^{tw_k}_s =  \frac{q^{\frac{s(s+3)}{2}}}{\{q\}^s}
\sum_{j=0}^s (-)^jq^{2j(j+1)k}\frac{[s]![2j+1]}{[s-j]![s+j+1]!} = \nn \\
= \frac{  q^{\frac{s(s+3)}{2}}}{\{q\}^s}
\sum_{j=0}^s [2j+1]e^{i\pi j}q^{2j(j+1)k}
\frac{q^{\frac{s(s+1)}{2}} \prod_{k=1}^s (1-q^{-2k}) }
{q^{\frac{(s-j)(s-j+1)}{2}}\prod_{k=1}^{s-j} (1-q^{-2k})
\cdot q^{\frac{(s+j+1)(s+j+2)}{2}}\prod_{k=1}^{s+j+1} (1-q^{-2k})} = \nn \\
= \{q\}
\sum_{j=0}^s [2j+1]e^{i\pi j}q^{2j(j+1)k} q^{-j^2-j-1}
 \frac{(q^{-2s},q^2)_\infty(1,q^2)_\infty }{ (q^{-2(s-j)},q^2)_\infty(q^{-2(s+j+1)},q^2)_\infty}
 \nn \\
 \longrightarrow
 \exp\left(\frac{1}{2\hbar}\left(\frac{2k-1}{2}(\log u)^2 +i\pi\log u
 + {\rm Li}_2(z^{-1})-{\rm Li}_2(uz^{-1}) - {\rm Li}_2(u^{-1}z^{-1}) + {\rm Li}_2(1)
 \right)+O(1)\right)
 \label{ac1}
\ee
This should be multiplied by
$\exp\left(\frac{1}{2\hbar}\left( 2\log\ft\cdot\log z +
  {\rm Li}_2(\ft^{-2}z^{-1})-{\rm Li}_2(\ft^{-2}z)
 \right)+O(1)\right) $
 from (\ref{S41}).

 The saddle point conditions $\p_z S=0$ and $\p_u S=0$ now are
 \be
\ft^2 (1-\ft^{-2}z_*)(1-\ft^{-2}z_*^{-1}) \cdot \frac{(1-z_*^{-1})}{(1-u_*z_*^{-1})(1-u_*^{-1}z_*^{-1})}=   1
\ee
and
\be
 e^{i\pi} u_*^{2k-1}\frac{1-u_*z_*^{-1}}{1-u_*^{-1}z_*^{-1}} = 1
\ \ \Longrightarrow \ \ z_* = \frac{u_*^{2k+1}+1}{u_*^{2k}+u_*}
\label{sp1}
\ee

Like in sec.\ref{QcL41} we have an alternative:
\be
a^{tw_k}_s =  \frac{q^{\frac{s(s+3)}{2}}}{\{q\}^s}
\sum_{j=0}^s (-)^jq^{2j(j+1)k}\frac{[s]![2j+1]}{[s-j]![s+j+1]!} = \nn \\
= \frac{  q^{\frac{s(s+3)}{2}}}{\{q\}^s}
\sum_{j=0}^s [2j+1]e^{i\pi j}q^{2j(j+1)k}
\frac{q^{-\frac{s(s+1)}{2}} \prod_{i=1}^s (q^{2i}-1) }
{q^{\frac{-(s-j)(s-j+1)}{2}}\prod_{i=1}^{s-j} (q^{2i}-1)
\cdot q^{\frac{-(s+j+1)(s+j+2)}{2}}\prod_{i=1}^{s+j+1} (q^{2i}-1)} = \nn \\
= \{q\}q^{s(s+3)}
\sum_{j=0}^s [2j+1]e^{i\pi (s+j+1)}q^{2j(j+1)k} q^{j^2+j+1}
 \frac{(q^{2s-2j},q^2)_\infty(q^{2s+2j+2},q^2)_\infty }{ (q^{2s},q^2)_\infty(1,q^2)_\infty}
 \nn \\
 \longrightarrow
 \exp\left(\frac{1}{2\hbar}\left(\frac{1}{2}(\log z)^2+\frac{2k+1}{2}(\log u)^2 +i\pi\log(uz)
 + {\rm Li}_2\left(\frac{z}{u}\right)+{\rm Li}_2(zu) - {\rm Li}_2(z)-{\rm Li}_2(1)
 \right)+O(1)\right)
 \label{ac2}
\ee

This should be multiplied by
$ \exp\left(\frac{1}{2\hbar}
\left({\rm Li}_2\left(\frac{\ft^2}{z}\right) - {\rm Li}_2(\ft^2 z)-2\log\ft \cdot\log z\right)\right)$
from (\ref{S41bb}).
Saddle point conditions are:
\be
z_*\ft^{-2}\left(1-\frac{\ft^2}{z_*}\right)(1-\ft^2z_*) = e^{i\pi}\frac{\left(1-\frac{z_*}{u_*}\right)(1-z_*u_*)}{1-z_*}
\ee
and
\be
e^{i\pi} u_*^{2k+1} = \frac{1-z_*u_*}{1-\frac{z_*}{u_*}}
\ \ \Longrightarrow \ \ z_* = \frac{u_*^{2k+1}+1}{u_*^{2k}+u_*}
\label{sp2}
\ee
At $\ft=1$  this gives:
\be
(z_*^2-2z_*+1)(z_*-1) = -(1+z_*^2)+z_*\left(u_*+\frac{1}{u_*}\right)
\label{st-1z}
\ee
and
\be
u_*^{4k}(u_*-1)-4u_*^{2k+1}-u_*^2+u_*=0
\label{ust}
\ee

The corresponding action is either
\be
2S(z_*,u_*) =
\left({\rm Li}_2\left(\frac{1}{z_*}\right) - {\rm Li}_2(z_*) \right)
\ \ \underline{-2i\pi \Big(k-\theta(k)+\theta(-k)\Big)\log (u_*)}
+ \nn \\
+ \left( \frac{2k-1}{2}(\log u_*)^2  + i\pi \log(u_* )
+ {\rm Li}_2\left(\frac{1}{z_*}\right) - {\rm Li}_2\left(\frac{u_*}{z_*}\right)+{\rm Li}_2\left(\frac{1}{z_*u_*}\right) +{\rm Li}_2(1)
 \right)
\label{acc1}
\ee
from (\ref{ac1}) or
\be
2S(z_*,u_*) =
\left({\rm Li}_2\left(\frac{1}{z_*}\right) - {\rm Li}_2(z_*) \right)
\ \ \underline{-2i\pi\log (z_*)-2i\pi\Big(k+2\theta(-k)-\delta_{k,-1}\Big)\log (u_*)}
+ \nn \\
+ \left(\frac{1}{2}(\log z_*)^2+\frac{2k+1}{2}(\log u_*)^2  + i\pi \log(u_*z_*)
 + {\rm Li}_2\left(\frac{z_*}{u_*}\right)+{\rm Li}_2(z_*u_*) - {\rm Li}_2(z_*)-{\rm Li}_2(1)
 \right)
\label{acc2}
\ee
from (\ref{ac2}).
The terms with $i\pi$ come from the sign factors in (\ref{ac1}) and (\ref{ac2}),
by the simple rules
 $(-)^s = e^{\i\pi s} = e^{\frac{i\pi\log z}{2\hbar}}$ and
$(-)^j = e^{\i\pi j} = e^{\frac{i\pi\log u}{2\hbar}}$.
However, they are defined modulo $2\pi$ and actually depend on the precise definition
of dilogarithmic function which is logarithmically ramified at $1$ and $\infty$.
By $\theta(k)$ we denote  the Heaviside function, which is $1$ for $k\geq 0$ and $0$ for $k<0$.
Matching between (\ref{ac1}) and (\ref{ac2}) is more or less guaranteed by (\ref{L+Linv}) --
modulo the phase ambiguity in logarithm is (\ref{LiL}).
One can see that the corresponding formulas (\ref{sp1}) and (\ref{sp2}) for $z_*$ are indeed the same.
However, there are many roots in the saddle point equations for $u_*$,
and they can match only after a careful adjustment of phases.

\subsubsection{Ambiguity and hypervolume}

Most important, different roots $u_*$ provide different values of the classical action,
as illustrated by the following list:

\be
\begin{array}{c|c|c|c|c}
k& {\rm knot} & 2\,{\rm Im}(S_*) \ {\rm from} \ (\ref{acc1}) & 2\,{\rm Im}(S_*) \ {\rm from} \ (\ref{acc2})
&{\rm hyperbolic\ volume}\\
\hline
-4 & 10_1 &  && {\bf 3.5262} \\
&&                   -6.086047830  & -4.364484056 \\
&&                    6.086047832 & 5.931259486 \\
&&                   -4.371453981 & -5.770152171 \\
&&                    4.371453978 & 4.371453979 \\
&&                   -3.692565502 & -7.227277662 \\
&&                    3.692565503 & 3.692565503 \\
&&                   -3.526195990 & -8.217587130 \\
\hline
&&                   {\bf 3.526195991}&  {\bf 3.526195990} \\
\hline
&&                    1.576932080& 2.133864626    \\
&&                   -1.576932079& -6.825255768  \\
&&                   -0.503795834& 0.695164548   \\
&&                    0.503795843& -4.229876697  \\
&&                   -2.920535197& -0.837109719  \\
&&                   2.920535197& -0.5615884729  \\
&&                    -5.964898570& -2.676559368 \\
&&                    5.964898574& 4.243334798
\end{array}
\ee

\be
\begin{array}{c|c|c|c|c}
k& {\rm knot} & 2\,{\rm Im}(S_*) \ {\rm from} \ (\ref{acc1}) & 2\,{\rm Im}(S_*) \ {\rm from} \ (\ref{acc2})
&{\rm hyperbolic\ volume}\\
\hline
-3 & 8_1 &&& {\bf 3.42721}   \\
&&                                 -5.197078051& -3.152892172\\
&&                     5.197078052& 4.815012509\\
&&                    -3.754596802& -6.053972813\\
&&                     3.754596806& 3.754596807\\
&&                    -3.427205246& -7.846147898\\
\hline  &&         {\bf 3.427205247}& {\bf 3.427205245}\\
\hline
&&                     1.014796782& 1.979760170\\
&&                    -1.014796781& -6.398702809\\
&&                    -1.738553398& 0.458706681\\
&&                    1.738553402& -2.758082687\\
&&                    -5.023851355& -1.317545136\\
&&                     5.023851360& 2.979665479\\
&&
\end{array}
\ee

\be
\begin{array}{c|c|c|c|c}
k& {\rm knot} & 2\,{\rm Im}(S_*) \ {\rm from} \ (\ref{acc1}) & 2\,{\rm Im}(S_*) \ {\rm from} \ (\ref{acc2})
&{\rm hyperbolic\ volume}\\
\hline
-2 & 6_1 &  & &                           {\bf  3.16396} \\
&&                   -3.960564138& -1.415104897\\
&&                    3.960564138& 2.902911933\\
&&                   -3.163963229& -6.829356351\\
\hline && {\bf 3.163963229} & {\bf 3.163963229}\\ \hline
&&                    0.047796851& 2.125241103\\
&&                   -0.047796851& -5.790634231\\
&&                   -3.675813582& 0.357452693\\
&&                    3.675813586& 1.130354343
\end{array}
\ee

\be
\begin{array}{c|c|c|c|c}
k& {\rm knot} & 2\,{\rm Im}(S_*) \ {\rm from} \ (\ref{acc1}) & 2\,{\rm Im}(S_*) \ {\rm from} \ (\ref{acc2})
&{\rm hyperbolic\ volume}\\
\hline
-1 & 4_1 &  & &                         {\bf 2.02988} \\
&&                   -2.029883213& -2.029883214\\
\hline &&                   {\bf 2.029883213}& {\bf 2.029883214}\\ \hline
&&                   -1.385248369& -1.385248368\\
&&                    1.385248369& 1.385248370
\end{array}
\ee

\be
\begin{array}{c|c|c|c|c}
k& {\rm knot} & 2\,{\rm Im}(S_*) \ {\rm from} \ (\ref{acc1}) & 2\,{\rm Im}(S_*) \ {\rm from} \ (\ref{acc2})
&{\rm hyperbolic\ volume}\\
\hline
2 & 5_2 &  && {\bf 2.82812}\\
&&                                  0.& -7.067316139\\
\hline && {\bf 2.828122089}& {\bf 2.828122090}  \\ \hline
&&                   -1.061293057& -1.167057066\\
&&                    3.533658068& 3.533658071\\
&&                   -2.828122089& -6.467544156\\
&&                   1.061293053& -2.472365018
\end{array}
\ee

\be
\begin{array}{c|c|c|c|c}
k& {\rm knot} & 2\,{\rm Im}(S_*) \ {\rm from} \ (\ref{acc1}) & 2\,{\rm Im}(S_*) \ {\rm from} \ (\ref{acc2})
&{\rm hyperbolic\ volume}\\
\hline
3 & 7_2 &  & &  {\bf  3.33174} \\
&&                    5.194119502& 5.194119501\\
&&                    -2.597059747& -7.791179248\\
&&                     3.843588654& 3.843588652\\
&&                    -3.843588652& -6.468665329\\
\hline &&  {\bf  3.331744231}& {\bf 3.331744230} \\ \hline
&&                    -3.331744230& -7.671305591\\
&&                    -1.285611129& -4.261083754\\
&&                    1.285611126& -0.078477613\\
&&                     1.045271020& 1.679434122 \\
&&                   -1.045271021& -4.304510803
\end{array}
\ee

\be
\begin{array}{c|c|c|c|c}
k& {\rm knot} & 2\,{\rm Im}(S_*) \ {\rm from} \ (\ref{acc1}) & 2\,{\rm Im}(S_*) \ {\rm from} \ (\ref{acc2})
&{\rm hyperbolic\ volume}\\
\hline
4 & 9_2 &  & &{\bf 3.48666}\\
&&                     6.260220410& 6.260220414\\
&&                    -4.173480274& -8.346960548\\
&&                     4.713033936& 4.713033933\\
&&                    -4.713033935& -6.766430657\\
&&                     3.712929489& 3.712929489\\
&&                    -3.712929490& -7.183050999\\
\hline &&            {\bf 3.486660145}& {\bf 3.486660144}\\ \hline
&&                    -3.486660147& -8.137019706\\
&&                    -1.688770925& -5.619974796\\
&&                    1.688770927& 0.969615238\\
&&                    0.245418035& -1.641364463\\
&&                    -0.245418042& -1.828757059\\
&&                    2.488023058& 3.315049131\\
&&                   -2.488023059& -5.368445855
\end{array}
\ee

\bigskip

The number of solutions for each knot is defined  by the order of (\ref{ust})
and is approximately $4|k|$.
The true hyperbolic volume
corresponds to just one of these many solutions.
Matching can be improved by more adequate selection of the action function
than implied by the naive (\ref{atwk}) --
but this corresponds to imposing additional restrictions on the definition
of the ``true'' Kashaev limit.
They can be natural from the point of view of various topological ideas,
but algebraic formulation in terms of $\Ap$-polynomials remains to be found.

Note that the underlined logarithmic terms in (\ref{acc1}) and (\ref{acc2}) 
are adjusted to  match the values at two columns at the ``proper'' points,
corresponding to the hyperbolic volume -- 
as we see from the tables, this does not mean full coincidence at other points.

\section{Quantum $A$-pols (equations w.r.t. $r$) and their Kashaev limits
\label{Apols}}

The differential equation, which a Jones polynomial satisfies as
a function of representation $r$, is known as {\it quantum $\Ap$-polynomial}
\cite{qAp,GMqap}.
Again it has two types of solutions -- leading to different Kashaev limits:
the naive $\nKL$ which is inverse Alexander
and the true $\KL$ with non-trivial classical action, which is reduces to hyperbolic volume
at particular point $\ft=1$ -- which, however, is a kind of an integration constant
for $\Ap$ equation and, strictly speaking, is not determined by it.
In between the two limits there is additional one -- to {\it classical} $A$-polynomial,
where the double scaling limit from $q$ to $\ft$ is already taken,
but $r$-shift operator $\hat L$ is assumed to act by multiplication $\hat L \rightarrow L$,
i.e. the limit $\nKL(J)$ is assumed to be an eigenfunction of $\hat L$:
\be
\Ap_q^\CK(\hat L, M) J = Q^\CK(M) \ \ \Longrightarrow \ \
\Ap_1^\CK(\hat L, M) J = Q^\CK(M) \ \ \Longrightarrow \ \ A^\CK(L,M)\KL(J) = Q^\CK (M)
\ee
In the theory of $A$-polynomials the variable $\ft$ is usually denoted by $M$.
The limit $\nKL$ assumes further that the eigenvalue $L=1$.
The Melvin-Morton conjecture implies then that the classical $A$-polynomial $A^{\cal K}(L,M)$
at $L=1$ gets proportional to the Alexander $\Al^\CK(\ft = M)$,
\be
A^\CK(1,M) = Q^\CK(M)\Al^\CK(M)
\ee

In this section we illustrate these steps with two simplest examples $\CK=3_1$ and $\CK=4_1$.
In the first (trefoil) case things are especially simple, because non-trivial limit $\KL$ does not exist --
the quasiclassical action always vanishes, and we remain with only the ''`naive'' limit $\nKL$.
Figure-eight is more representative, it is a hyperbolic knot, where both $\KL$ and $\nKL$ make sense.
Still, it is also an over-simplification, because quasiclassics is not truly branched,
the limit $\nKL$ and hyperbolic volume are defined unambiguously -- in any way one prefers.
This is basically because $\Cp$-polynomial is trivial in this case, in other examples in the next sections
the story will get far less transparent.

\subsection{Example of $3_1$}

The colored Jones polynomial $J^{3_1}_r$ for the trefoil satisfies
\be
\hat \Ap^{3_1}_q J_r :=
q^{6r+2}(q^{2r+2}-1)J^{3_1}_r +  (q^{2r}-1)J^{3_1}_{r-1} = q^{2r+2}(q^{4r+2}-1)
\label{eq31}
\ee
Define an operator $J_{r-1}=\hat L J_r$.
Then if we denote multiplication by $\ft=q^r$ through $M$, and put all the remaining $q=1$,
we get from (\ref{eq31})
\be
\hat \Ap^{3_1}_1(\hat L,M) \, {\cal J}^{3_1} := (\hat L + M^6){\cal J}^{3_1} =
\underbrace{ M^2(M^2+1)}_{Q^{3_1}(M)}
\label{LMeq31}
\ee
We can now consider two different cases.
One is to assume that the limiting ${\cal J}$, where only $\ft$-dependence is left,
while otherwise $q=1$, is an eigenfunction of the $r$-shift operator $\hat L$
so that the l.h.s. becomes
\be
A^{3_1}(L,M)   = L + M^6    
\label{clAp31}
\ee
This is the standard expression for classical $A$-polynomials of $\CK=3_1$, see \cite{katlas}.
However, it is unclear what the inhomogeneous classical equation with $Q^{3_1}(M)$ at the r.h.s.
would mean.
Instead classical $A$-polynomial has a variety of alternative interpretations --
not referring to quantum $\Ap$-polynomials, and allowing its algorithmic evaluation. 

Another, stronger but clearer assumption is that ${\cal J}$  is fully independent of $r$,
i.e. $\hat L $, acts as unity, what can be interpreted in terms of (\ref{clAp31}) as putting $L=1$.
We call such better grounded approximation to ${\cal J}$ the $\nKL$-limit, $\nKL(J)$.
It satisfies
\be
\nKL(J^{3_1}) = \frac{\mathfrak{t}^2(\mathfrak{t}^4-1)}{(\mathfrak{t}^2-1)(\mathfrak{t}^6+1)}
= \frac{1}{\mathfrak{t}^2+\mathfrak{t}^{-2}-1} = \frac{1}{{\Al}^{3_1}(\mathfrak{t})}
\label{nKL31}
\ee
It looks exactly the same in terms of $M$:
\be
\nKL(J^{3_1}) = \frac{M^2(M^2+1)}{A^{3_1}(1,M)} =  \frac{M^2(M^2+1)}{1+M^6} = \frac{1}{\Al^{3_1}(M)}
\ee
Thus we see that the Melvin-Morton conjecture implies that the classical $A$-polynomial at $L=1$,
$A^\CK(1,M)$ is proportional to Alexander $\Al^\CK(M)$.
This, however, assumes that the $A$-polynomial is defined/normalized so that the free term $Q^\CK(M)$
in (\ref{LMeq31}) is polynomial rather than rational.
Such requirement is not always fulfilled for classical $A$-polynomials in \cite{katlas} --
because the control of free terms actually requires the knowledge of  {\it quantum} $\Ap$-polynomials --
which is far more complicated than just the knowledge of homogeneous part of classical ones.
We will see in sec.\ref{clAp} below that indeed for some knots $\CK$  some factors in $\Al^\CK(M)$
are not present in conventional  $A^\CK(1,M)$ from \cite{katlas}.

The free term $Q^{\cal K}$ is also neglected in the ``true'' Kashaev limit $\KL(J)$,
where it is  assumed that ${\cal J}$ is "large" -- behaves as $\exp(S/\hbar)$ with non-vanishing $S$.

However, in the case of $\CK=3_1$ a solution of the homogeneous equation (with the r.h.s. omitted) is
\be
\frac{(-)^r q^{-3Q^2-5r}}{q^{2r}-1}
\label{clasol31}
\ee
and the assumption is not self-consistent,
because (\ref{clasol31}) is small rather than large and can not justify the
omission of the free term at the r.h.s. of (\ref{eq31}). 
Thus in the case of trefoil (\ref{nKL31}) is the only sensible quasiclassical limit.
This is in a nice accordance with the vanishing of hyperbolic volume for the trefoil. 

\subsection{Example of $4_1$}

To get  two distinct limits we need to consider at least the figure eight knot $4_1$. 
This time the equation is of the second order:
\be
q^{2r}\Big((q^{4r-2}-1)(q^{2r+2}-1)J^{4_1}_r + (q^{4r+2}-1)(q^{2r-2}-1)J^{4_1}_{r-2}\Big) - \nn \\
- (q^{4r}-1)(q^{2r}-1)(q^{6r}-q^{4r}-q^{2r+2}-q^{2r-2}-1+q^{-2r})J^{4_1}_{r-1} = (q^{2r}+1)(q^{4r+2}-1)(q^{4r-2}-1)
\label{eq41}
\ee
or
\be
\!\!
q^{4r}\left(\frac{q^{2r+2}-1}{q^{4r+2}-1}J^{4_1}_r + \frac{q^{2r-2}-1}{q^{4r-2}-1}J^{4_1}_{r-2}\!\right)
- (q^{4r}-1)(q^{2r}-1)\!\left(1-\frac{q^{2r}(q^{4r}+1)}{(q^{4r+2}-1)(q^{4r-2}-1)}\!\right)\!J^{4_1}_{r-1}\!
=\!q^{2r}(q^{2r}+1)
\ee
Substituting $q^r=\mathfrak{t}$ and
\be
J(r-1) = e^{\frac{ S(\mathfrak{h})}{\bar h}+S'(\mathfrak{h})+\ldots}
= {\cal J}e^{S'} + O(\hbar)
\ee
we get
\vspace{-0.7cm}
\be
\underline{\overbrace{\left(e^{S'}-2+e^{-S'}\right)}^{
\left(e^{\frac{1}{2}S^\prime}-e^{-\frac{1}{2}S^\prime}\right)^2}{\cal J}}
+(\mathfrak{t}+\mathfrak{t}^{-1})^2
\Big(\overbrace{(3-\mathfrak{t}^2-\mathfrak{t}^{-2})}^{{\rm Al}^{4_1}(\mathfrak{t})}{\cal J}
-\boxed{1}\Big) = O(\hbar)
\label{EQ41}
\ee
Kashaev limit corresponds to neglecting all positive powers of $\hbar$, i.e. putting the r.h.s. to zero.
Still after that we have two options:
the ``naive''  limit, when $S=0$ and underlined term can be neglected:
\be
\nKL(J^{4_1})
= \frac{\boxed{1}}{3-\mathfrak{t}^2-\mathfrak{t}^{-2}} = \frac{1}{{\rm Al}^{4_1}(\mathfrak{t})}
\ee
and the ``true'' $\KL$, when neglected is the free term in the box,
equation becomes homogeneous and $S_{\KL}$ is an integral of its $\mathfrak{t}$-derivative,
which satisfies
$\left(e^{\frac{1}{2}S^\prime}-e^{-\frac{1}{2}S^\prime}\right)^2=G^2
=-(\mathfrak{t}+\mathfrak{t}^{-1})^2 {{\rm Al}^{4_1}(\mathfrak{t})}
= -4-\mathfrak{t}^2-\mathfrak{t}^{-2}+\mathfrak{t}^4+\mathfrak{t}^{-4}$
or
$e^{\frac{1}{2}S^\prime}-e^{-\frac{1}{2}S^\prime}=\pm G$:
\be
S' = \frac{\p S}{\p \mathfrak{h}} =\mathfrak{t} \frac{\p S}{\p \mathfrak{t}} =
2\,\log\left(\frac{G\pm \sqrt{G^2+4}}{2}\right)
= 2\,\log\left(\frac{\sqrt{4+\mathfrak{t}^2+\mathfrak{t}^{-2}-\mathfrak{t}^4-\mathfrak{t}^{-4}}
\pm \sqrt{\mathfrak{t}^2+\mathfrak{t}^{-2}-\mathfrak{t}^4-\mathfrak{t}^{-4}}}{2i}\right),
\label{S'*}
\ee
Thus
\be
S_{\KL}= \int  2\,\log\left(\frac{\sqrt{4+\mathfrak{t}^2+\mathfrak{t}^{-2}-\mathfrak{t}^4-\mathfrak{t}^{-4}}
\pm \sqrt{8+\mathfrak{t}^2+\mathfrak{t}^{-2}-\mathfrak{t}^4-\mathfrak{t}^{-4}}}{2i}\right)
\frac{d\mathfrak{t}}{\mathfrak{t}} + {\rm const} 
\label{SKL41}
\ee
In variance with the case of  (\ref{geq41a}) we deal here with $e^{\frac{1}{2}S^\prime}$ rather than with $g=e^{S'}$.
Thus the quadratic equations are different:
\be
e^{\frac{1}{2}S^\prime}-e^{-\frac{1}{2}S^\prime} = G(t)  \Longrightarrow e^{\frac{1}{2}S^\prime} = \frac{G\pm\sqrt{G^2+4}}{2}
\Longrightarrow e^{S^\prime} = \frac{G^2+2\pm\sqrt{(G^2(G^2+4)}}{2}
\Longrightarrow e^{S^\prime} -(G^2+2) + e^{-S^\prime} = 0
\nn
\ee
with $-(G^2+2) = 2+\mathfrak{t}^2+\mathfrak{t}^{-2}-\mathfrak{t}^4-\mathfrak{t}^{-4}$,
in full accordance with (\ref{geq41a}) where
$g(\mathfrak{t})=e^{S^\prime(\mathfrak{t})} = e^{\mathfrak{t}\,\p_\mathfrak{t} S}$.

Note once again that the $\ft$-independent part of the action, i.e. the hyperbolic volume (\ref{hypvol41})
drops out of the underlined part of the equation and thus remains undefined in (\ref{SKL41}).

From (\ref{eq41}) we can extract the standard quantum and classical $A$-polynomials,
by substituting $M=\ft=q^r$ and eliminating all other $q=1$:
\be
\hat\Ap^{4_1}_1 {\cal J}^{4_1} :=
M^4(\hat L^2 {\cal J}^{4_1}+{\cal J}^{4_1})-(M^8-M^6-2M^4-M^2+1)\hat L{\cal J}^{4_1}
= \underbrace{M^2(M^2+1)^2}_{Q^{4_1}(M)}
\label{LMeq41} \\
\Longrightarrow \ \
A^{4_1}(L,M) = L^2M^4 - (M^8-M^6-2M^4-M^2+1)L + M^4
\nn
\ee
At $L=1$ the classical $A$-polynomial is divisible by Alexander:
\be
A^{4_1}(1,M)
- M - 1)(M^2 + 1)^2 =  Q^{4_1}(M)\cdot \Al^{4_1}(M)
\ee
and the extra factor is exactly the free term $Q^{4_1}(M)$ in (\ref{LMeq41}).

\subsection{The main claim(s)
\label{mainclaim}}

In general the non-homogeneous  $\Ap$-polynomial equation looks like
\be
\boxed{
\underline{P_{\cal K} (e^{S'}) {\cal J}^{\cal K}}
-\Big(Al^{\cal K}(\mathfrak{t}){\cal J}^{\cal K} -\boxed{1}\Big) Q_{\cal K}(\mathfrak{t}) = O(\hbar)
}
\label{ApolvsAl}
\ee
with some knot-dependent polynomials $P$ and $Q$, such that $P(1)=0$.
It has two limits when $\hbar\longrightarrow 0$.
One, $\nKL(J)$ is inverse Alexander with $S=0$, $P_{\cal K} (S')=P_{\cal K} (1)=0$, and underlined term eliminated.
Another is more sophisticated {\it true}  $\KL(J)$ -- when neglected is the boxed free term without ${\cal J}$,
but $P$ is non-trivial.
Then ${\cal J}$ can be omitted and $S$ defined a  solution of an elementary differential equation.
If we add the relation of equation (\ref{ApolvsAl}) to classical $A$-polynomial $A^{\CK}(L,M)$,
then we obtain from (\ref{ApolvsAl}) that
\be
\ \ \ \ \ \ \ \ \ \ \ \ \ \ \ \ \ \ \ \ \ \ \ \ \ \ \
\boxed{
A^\CK(L=1,M)\ \vdots\ \Al^\CK(M)
}
\ \ \ \ {\rm for\ appropiately\ defined} \ A^\CK
\label{AldividesAcl}
\ee
This is the second part of our main claim.
As already mentioned, it applies to the properly defined $A$-polynomial, which comes from the quantum
equation (\ref{ApolvsAl}) with the {\it polynomial} free term $Q_\CK(\ft)$.

The fact that Alexander appears in these equations is somewhat surprising, but it seems to be a fact.
Actually, the relevance of Alexander for $A$-polynomials was already noted in \cite{Cooper}.
Moreover, according to \cite{KLM}, this should have a direct generalization to arbitrary knot polynomials
with $N>2$, but analysis requires better technique for finding {\it colored} quantum $\Ap$-polynomials
and will be presented elsewhere, see \cite{GMqap}.

\section{A few  examples of $\Ap$-polynomial decomposition \label{exaAp}}

In this section we list a few examples of (\ref{ApolvsAl}) for the simplest knots.
This is the list which deserves further extension and, hopefully, things will always work this way.

\subsection{$\CK = [3,1]$}

\be
{\rm qAp}: \ \ \ \  \ft ^6 q^2(\ft^2 q^2-1)J_r +(\ft^2-1)J_{r-1} = \ft^2q^2(\ft^4q^2-1) \nn
\ee
\be
\downarrow q=1 \ \ \ \ \ \ \ \ \ \ \ \ \nn
\ee
\be
\ft ^6  (\ft^2  -1){\cal J}_r +(\ft^2-1){\cal J}_{r-1} = \ft^2 (\ft^4 -1) \nn
\ee
\be
\updownarrow \ \ \ \ \ \ \ \ \ \ \ \ \nn
\ee
\be
{\rm KLAp}: \ \ \ \ \boxed{
\underline{\ft^6({\cal J}_r-{\cal J}_{r-1]})} +
\underbrace{\underline{(\ft^6+1)\Big({\cal J}_{r-1}} - \frac{1}{\ft^2 - 1 + \ft^2}\Big)}
} \nn
\ee
\be
\begin{array}{ccc}
\KL \swarrow && \searrow \nKL \\
\\
\ft^6({\cal J}_r-{\cal J}_{r-1}) + (\ft^6+1){\cal J}_{r-1} = 0
&\ \ \ \ \ \ \ \  \ \ \ \ \ \ \ \ \ \ \ \ \ \ \ \  &
{\cal J}_{r-1} = \frac{1}{\ft^2 - 1 + \ft^{-2}}  \\ \\
\updownarrow && \updownarrow \\ \\
e^{S'}-1 + 1+\ft^{-6}= 0 && \nKL(J) = \frac{1}{\Al(\ft)}
\\ \\ \downarrow \\ \\
S^{[3,1]}(\ft) = -3 (\log\ft)^2 + 6i\pi\log\ft
\\ \\ \downarrow \\ \\
 S^{[3,1]}(\ft=1) = 0
\end{array}
\ee
Standing in the box is our main qAp equation (\ref{main2})=(\ref{ApolvsAl}) --
the specially rewritten double-scaling limit of quantum $\Ap$-polynomial equation.
Underlined are the terms, which form homogeneous equation, and lead to the
standard Kashaev limit with an equation for $(\log\ft)$-derivative $S'=\ft \,dS/d\ft $ of
the "classical action" -- which can be further integrated with
the hyperbolic volume of $S^3/\CK$ as the ``initial condition''
at $\ft =1$ (which actually vanishes for the trefoil $\CK=3_1$,
which is a torus rather then hyperbolic knot).
Underbraced are the terms, relevant in naive Kashaev limit,
when the action is zero, and the answer is actually the inverse Alexander.

\subsection{$\CK = [4,1]$}

\be
{\rm qAp}: \ \ \ \  (\ft^4 q^{-2}-1)(\ft^2 q^2-1)J_r
+ (\ft^4q^2-1)(\ft^2q^{-2}-1)J_{r-2} - \nn \\
- (\ft^4-1)(\ft^2-1)(\ft^4-\ft^2-q^-q^{-2}-\ft^{-2}+\ft^{-4})J_{r-1}
= \ft^{-2}(\ft^2+1)(\ft^4q^2-1)(\ft^4q^{-2}-1) \nn
\ee
\be
\downarrow q=1 \ \ \ \ \ \ \ \ \ \ \ \ \nn
\ee
\be
(\ft^4  -1)(\ft^2  -1){\cal J}_r
+ (\ft^4 -1)(\ft^2 -1){\cal J}_{r-2}
- (\ft^4-1)(\ft^2-1)(\ft^4-\ft^2-2-\ft^{-2}+\ft^{-4}){\cal J}_{r-1}
= \ft^{-2}(\ft^2+1)(\ft^4 -1)^2 \nn
\ee
\be
\updownarrow \ \ \ \ \ \ \ \ \ \ \ \ \nn
\ee
\be
{\rm KLAp}: \ \ \ \ \boxed{
\underline{ ({\cal J}_r-2{\cal J}_{r-1]}+{\cal J}_{r-2})} -
\underbrace{\underline{(\ft^4-\ft^2-4-\ft^{-2}+\ft^{-4})\Big({\cal J}_{r-1}} - \frac{1}{3-\ft^2 - \ft^{-2}}\Big)}
} \nn
\ee
\be
\begin{array}{ccc}
\KL \swarrow && \searrow \nKL \\
\\
({\cal J}_r-2{\cal J}_{r-1]}+{\cal J}_{r-2}) = (\ft^4-\ft^2-4-\ft^{-2}+\ft^{-4}) {\cal J}_{r-1}
&\ \ \ \ \ \ \ \  \ \ \ \ \ \ \ \ \ \ \ \ \ \ \ \  &
{\cal J}_{r-1} = \frac{1}{3-\ft^2 - \ft^{-2}}  \\ \\
\updownarrow && \updownarrow \\ \\
\Big(e^{S'/2}-e^{-S'/2}\Big)^2 = (\ft^4-\ft^2-4-\ft^{-2}+\ft^{-4}) && \nKL(J) = \frac{1}{\Al(\ft)}
\\ \\
\downarrow \\ \\
(\ref{SKL41}) \\ \\
\downarrow  \\ \\
(\ref{hypvol41})
\end{array}
\ee

\subsection{$\CK=[5,1]$}

\be
{\rm qAp}: \ \ \ \ \ft^{10} q^{4} (\ft^2  q^2 -1) J_r
+  (\ft^2-1) J_{r-1}
= \ft^{4}q^{4}(\ft^4q^2-1)  \nn
\ee
\be
\downarrow q=1 \ \ \ \ \ \ \ \ \ \ \ \ \nn
\ee
\be
\ft^{10} (\ft^2   -1) {\cal J}_r
+    (\ft^2-1) {\cal J}_{r-1}
= \ft^{4}(\ft^4 -1)
\nn
\ee
\be
\updownarrow \ \ \ \ \ \ \ \ \ \ \ \ \nn
\ee
\be
{\rm KLAp}: \ \ \ \ \boxed{
\underline{ ({\cal J}_r - {\cal J}_{r-1} ) } +
\underbrace{\underline{(1+\ft^{-10} ) \Big({\cal J}_{r-1}} - \frac{1}{\ft^4-\ft^2 +1- \ft^{-2}+\ft^{-4}}\Big)}
} \nn
\ee
\be
\begin{array}{ccc}
\KL \swarrow && \searrow \nKL \\
\\
({\cal J}_r - {\cal J}_{r-1} ) +(1+\ft^{-10} ){\cal J}_{r-1}=0
&\ \ \ \ \ \ \ \  \ \ \ \ \ \ \ \ \ \ \ \ \ \ \ \  &
{\cal J}_{r-1} = \frac{1}{\ft^4-\ft^2 +1- \ft^{-2}+\ft^{-4}}
\\ \\
\updownarrow && \updownarrow \\ \\
e^{S'} +\ft^{-10}=0 && \nKL(J) = \frac{1}{\Al(\ft)}
\\ \\
\downarrow \\ \\
S^{5_1}(\ft) = - 5(\log\ft)^2 + 10i\pi\log\ft    \\ \\
\downarrow   \\ \\
S^{5_1}(\ft=1) = 0
\end{array}
\ee

\subsection{$\CK = [5,2]$}

The quantum A-polynomial relation in this case is \cite{GK}
\be
(\ft^4q^6-1)(\ft^2q^4-1)(\ft^2q^4+1)(\ft^2q^8-1)J_{r+3}
+ \nn \\
+ q^2(\ft^4q^6-1)(\ft^2q^6-1)^2(\ft^2q^6+1)\Big(\ft^{10}q^{28}-\ft^8q^{22}-\ft^6(q^{20}-q^{18}-q^{16}+q^{14})
+ \ft^4(q^{14}+q^8)+2\ft^2q^6-1\Big) J_{[r+2]}
-\nn \\
-\ft^4q^{14}(\ft^4q^{14}-1)(\ft^2q^4-1)^2(\ft^2q^4+1)
\Big(\ft^{10}q^{22}-2\ft^8q^{18}-\ft^6(q^{16}+q^{10})+\ft^4(q^{12}-q^{10}-q^8+q^6)+\ft^2q^4-1\Big) J_{r+1} +
\nn \\
+ \ft^{14}q^{32}(\ft^4 q^{14}-1)(\ft^2q^2-1)(\ft^2q^6-1))\ft^2q^6+1)J_r
\ =\ \ft^4q^{12}(\ft^4q^6-1)(\ft^4q^{10}-1)(\ft^4q^{14}-1)(\ft^2q^6+1)(\ft^2q^4+1)
\nn
\ee
\be
\downarrow q=1
\nn
\ee
\be
{\cal J}_{r+3}+(\ft^{10}-\ft^8+2\ft^4+2\ft^2-1){\cal J}_{r+2} - \ft^4(\ft^{10}-2\ft^8-2\ft^6+\ft^2-1){\cal J}_{r+1}
+\ft^{14}{\cal J}_r = \ft^4(\ft^2+1)^3
\ee
\be
\updownarrow
\nn
\ee
\be
\underline{({\cal J}_{r+3}-{\cal J}_r)+(\ft^{10}-\ft^8+2\ft^4+2\ft^2-1)({\cal J}_{r+2}-{\cal J}_r) - \ft^4(\ft^{10}-2\ft^8-2\ft^6+\ft^2-1)({\cal J}_{r+1}-{\cal J}_r)
+} \nn \\
+\underbrace{\underline{\Big(1+(\ft^{10}-\ft^8+2\ft^4+2\ft^2-1)  - \ft^4(\ft^{10}-2\ft^8-2\ft^6+\ft^2-1) +\ft^{14}\Big)
\Big({\cal J}_r} -\frac{1}{\Al(\ft)}\Big) \underline{\phantom{a_{\int_{\int b}}}\!\!\!\!\!\!\!\!\!\! = 0} }
\ee

The underbraced relation is true because
\be
\frac{1+(\ft^{10}-\ft^8+2\ft^4+2\ft^2-1)  - \ft^4(\ft^{10}-2\ft^8-2\ft^6+\ft^2-1) +\ft^{14}}{\ft^4(\ft^2+1)^3}
= \Al^{5_2}(\ft)
\ee

Underlined equation implies
\be
(e^{3S'}-1)+(\ft^{10}-\ft^8+2\ft^4+2\ft^2-1)(e^{2S'}-1) - \ft^4(\ft^{10}-2\ft^8-2\ft^6+\ft^2-1)(e^{S'}-1)
= \nn \\
= 1+(\ft^{10}-\ft^8+2\ft^4+2\ft^2-1)  - \ft^4(\ft^{10}-2\ft^8-2\ft^6+\ft^2-1) +\ft^{14}
\ee
to which we should add that   $S(\ft=1)={\rm hypvolume}(S^3/5_2)$.

\subsection{Arbitrary two-strand knot $\CK = [2m+1,1]$}

\be
{\rm qAp}: \ \ \ \ \ft^{4m+2}q^{2m} (\ft^2  q^2 -1) J_r
+  (\ft^2-1) J_{r-1}
= \ft^{2m}q^{2m} (\ft^4q^2-1)  \nn
\ee
\be
\downarrow q=1 \ \ \ \ \ \ \ \ \ \ \ \ \nn
\ee
\be
 \ft^{4m+2}(\ft^2   -1) {\cal J}_r
+   (\ft^2-1) {\cal J}_{r-1}
= \ft^{2m}(\ft^4 -1)
\nn
\ee
\be
\updownarrow \ \ \ \ \ \ \ \ \ \ \ \ \nn
\ee
\be
{\rm KLAp}: \ \ \ \ \boxed{
\underline{ ({\cal J}_r - {\cal J}_{r-1} ) } +
\underbrace{\underline{(\ft^{-4m-2}+1) \Big({\cal J}_{r-1}} - \frac{\ft^{2m}(\ft^2+1)}{\ft^{4m+2}+1}\Big)}
} \nn
\ee
\be
\begin{array}{ccc}
\KL \swarrow && \searrow \nKL \\
\\
({\cal J}_r - {\cal J}_{r-1} ) +(\ft^{-4m-2}+1){\cal J}_{r-1}=0
&\ \ \ \ \ \ \ \  \ \ \ \ \ \ \ \ \ \ \ \ \ \ \ \  &
{\cal J}_{r-1} = \left(\frac{\ft^{2m+1}+\ft^{-2m-1}}{\ft+\ft^{-1}}\right)^{-1}
\\ \\
\downarrow && \downarrow \\ \\
e^{S'} +\ft^{-4m-2}=0 && \nKL(J) = \frac{1}{\Al(\ft)}
\\ \\
\downarrow \\ \\
S^{2m+1_1}(\ft) = -(2m+1)(\log\ft)^2 + (4m+2)i\pi\log\ft    \\ \\
\downarrow   \\ \\
S^{5_1}(\ft=1) = 0
\end{array}
\ee

\section{Alexander within classical $A$-polynomials
\label{clAp}}

In this section we provide some examples of (\ref{AldividesAcl}).
They will also demonstrate the reservations about this statement for
the standard classical $A$-polynomials from \cite{katlas}.

The simplest from this perspective are torus knots, which have nearly trivial
classical $A$-polynomials:
\be
 A^{{\rm tor}[m,n]} = L+M^{mn}
\ee
For two-strand torus knots Alexander is nearly the same:
\be
\Al^{{\rm tor}[2,2k+1]}(M) =  \frac{M^{2(2k+1)}+1}{M^{2k}(M^2+1)}
\ \ \Longleftrightarrow \ \   A^{{\rm tor}[m,2k+1]} = M^{2k}(M^2+1)\Al^{{\rm tor}[2,2k+1]}
\ee
However already for ${\rm tor}[3,4]$ exact factorization fails:
\be
\Al^{{\rm tor}[3,4]}(M) = (M^2-1+M^{-2})(M^4-1+M^{-4}) \nn \\
\ \ \Longleftrightarrow \ \   A^{{\rm tor}[3,4]}
= \frac{ M^6(M^2+M^{-2})( M+M^{-1})}  {M^3+M^{-3}}
\Al^{{\rm tor}[3,4]}(M)
\ee
with non-trivial denominator.

\bigskip

The next are twist knots with $\Al^{{\rm tw}_k} = -(2k-1)+k(M^2+M^{-2})$.
Their classical $A$-polynomials are already more sophisticated:

{\footnotesize
\be
A^{{\rm tw}_{1}}=  A^{3_1} = L+M^6, \nn \\ \nn \\
A^{{\rm tw}_{-1}}= A^{4_1} =L^2M^4 + L(-M^8 + M^6 + 2M^4 + M^2 - 1) + M^4
\nn \\ \nn \\
A^{{\rm tw}_{2}}=A^{5_{2}} = L^3 M^{14}+L^2 \left(-M^{14}+2 M^{12}+2 M^{10}-M^6+M^4\right)+L \left(M^{10}-M^8+2 M^4+2 M^2-1\right)+1
\nn \\ \nn \\
A^{{\rm tw}_{-3}}= A^{6_{1}} = L^4 M^8+L^3 \left(-2 M^{12}+3 M^{10}+3 M^8+M^2-1\right)+\nn \\
+ L^2 \left(M^{16}-3 M^{14}-M^{12}+3 M^{10}+6 M^8+3 M^6-M^4-3 M^2+1\right)+L \left(-M^{16}+M^{14}+3 M^8+3 M^6-2 M^4\right)+M^8
\nn \\ \nn \\
A^{{\rm tw}_{3}}=
A^{7_{2}} =
L^5+L^4 \left(M^{14}-M^{12}+3 M^4+4 M^2-2\right)+ \nn \\
+ L^3 \left(-2 M^{18}+5 M^{16}+M^{14}-4 M^{12}+6 M^8+5 M^6+2 M^4-4 M^2+1\right)+\nn \\
+ L^2 \left(M^{22}-4 M^{20}+2 M^{18}+5 M^{16}+6 M^{14}-4 M^{10}+M^8+5 M^6-2 M^4\right)+L \left(-2 M^{22}+4 M^{20}+3 M^{18}-M^{10}+M^8\right)+M^{22} \nn \\ \nn \\
\ldots
\nn
\ee
}
Still at $L=1$ they simplify to
\be
A^{{\rm tw}_{k}}(1,M) =  M^{2|k|}(M^2+1)^{(2k-1)\theta(k)+2k\theta(-k)}\cdot \Al^{{\rm tw}_{k}}(M)
\ee

Two other knots from the beginning on the Rolfsen table

{\footnotesize
\be
A^{6_{2}} = L^5 M^{26}+L^4 \left(-M^{30}+2 M^{28}-M^{26}-2 M^{24}+5 M^{22}+5 M^{20}-3 M^{18}\right)+\nn \\
+ L^3 \left(-M^{28}+3 M^{26}-M^{24}-5 M^{22}-3 M^{20}+12 M^{18}+13 M^{16}-3 M^{14}-8 M^{12}+3 M^{10}\right)+\nn \\
+ L^2 \left(3 M^{20}-8 M^{18}-3 M^{16}+13 M^{14}+12 M^{12}-3 M^{10}-5 M^8-M^6+3 M^4-M^2\right)+\nn \\
+L \left(-3 M^{12}+5 M^{10}+5 M^8-2 M^6-M^4+2 M^2-1\right)+M^4
\nn \\
\nn \\
A^{6_{3}} = L^6 M^{14}+L^5 \left(2 M^{20}-5 M^{18}+M^{16}+10 M^{14}+M^{12}-5 M^{10}+2 M^8\right)+\nn \\
+ L^4 \left(M^{26}-4 M^{24}+4 M^{22}+2 M^{20}-6 M^{18}+2 M^{16}+17 M^{14}+2 M^{12}-6 M^{10}+2 M^8+4 M^6-4 M^4+M^2\right)+\nn \\
+ L^3 \left(M^{28}-5 M^{26}+3 M^{24}+9 M^{22}-2 M^{20}-21 M^{18}+8 M^{16}+34 M^{14}+8 M^{12}-21 M^{10}-2 M^8+9 M^6+3 M^4-5 M^2+1\right)+\nn \\
+ L^2 \left(M^{26}-4 M^{24}+4 M^{22}+2 M^{20}-6 M^{18}+2 M^{16}+17 M^{14}+2 M^{12}-6 M^{10}+2 M^8+4 M^6-4 M^4+M^2\right)+\nn \\
+L \left(2 M^{20}-5 M^{18}+M^{16}+10 M^{14}+M^{12}-5 M^{10}+2 M^8\right)+M^{14}
\nn
\ee
}
also demonstrate full Alexander divisibility at $L=1$:
\be
A^{6_{2}}(1,M) = M^4(M^{12} - 3M^{10} + 5M^8 - 5M^6 + 5M^4 - 3M^2 + 1)(M^2 + 1)^5\cdot \Al^{6_2}(M)
\ee
with $\Al^{6_2}(M) = -\frac{M^8 - 3*M^6 + 3*M^4 - 3*M^2 + 1}{M^4}$
and
\be
A^{6_{3}}(1,M) =M^4(M^2 + M - 1)^2(M^2 - M - 1)^2(M^2 + 1)^6 \cdot \Al^{6_3}(M)
\ee
with $ \Al^{6_2}(M) = \frac{(M^4 - M^3 - M^2 + M + 1)(M^4 + M^3 - M^2 - M + 1)}{M^4}$.

\bigskip

Among non-fully divisible at $L=1$ are $A^{8_{18}}$ and $A^{8_{21}}$:

{\footnotesize
\be
A^{8_{18}}(1,M) = -\frac{M^6(M^8 - M^6 - 2M^4 - M^2 + 1)^2(M^4 - 4M^2 + 1)^2(M^{12} - 3M^{10} + 2M^8 + M^6 + 2M^4 - 3M^2 + 1)^2(M^2 + 1)^4}{(M^2 + M - 1)(M^2 - M - 1)}\cdot \Al^{8_{18}}
\nn \\ \nn \\
A^{8_{21}}(1,M) = \frac{M^4(M^4 + 1)(M^8 - 5M^6 + 10M^4 - 5M^2 + 1)(M^2 + 1)^3}{M^4 - M^2 + 1} \cdot \Al^{8_{21}}
\nn
\ee
}
with $\Al^{8_{18}} = -\frac{(M^2 + M - 1)(M^2 - M - 1)(M^4 - M^2 + 1)^2}{M^6}$
and $\Al^{8_{21}}  = -\frac{ (M^2 + M - 1)(M^2 - M - 1)(M^4 - M^2 + 1)}{M^4}$.

\bigskip

Rather complicated is the knot $9_{24}$, where classical $A$-polynomial has degree 19 in $L$.
Still at $L=1$ it simplifies and  acquires
$$\Al^{9_{24}}(M) = -\frac{(M^2 + M - 1)(M^2 - M - 1)(M^4 - M^2 + 1)^2}{M^6}$$
as a simple factor:
\be
A^{9_{24}}(1,M) = 4\Al^{9_{24}}(M) \, M^8 (M^2 + 1)^{17}
\Big(M^{64} - 27M^{62} + 347M^{60} - 2820M^{58} + 16239M^{56} - 70365M^{54}+ \nn\\
 + 237751M^{52} - 640319M^{50} + 1392506M^{48} - 2459518M^{46} + 3523630M^{44}  - 4049352M^{42} + 3618611M^{40} - \nn \\ 2290768M^{38} + 620455M^{36} + 697305M^{34} - 1187351M^{32} + 697305M^{30} + 620455M^{28} - 2290768M^{26} +\nn \\
 + 3618611M^{24} -   4049352M^{22} + 3523630M^{20} - 2459518M^{18} + 1392506M^{16} - 640319M^{14}
+ \nn \\
 + 237751M^{12} - 70365*M^{10} + 16239M^8 - 2820M^6 + 347M^4 - 27M^2 + 1\Big) \ \ \ \ \ \ \
\ee

One can easily get more examples with the help of explicit expressions for classical $A$-polynomials
at \cite{katlas}.

\section{$\Cp$-polynomials
\label{Cpols}}

Instead of writing equations for the entire knot polynomial,
one can impose them on the coefficients of its differential expansion \cite{DE}.
These equations are known a {\it quantum $C$-polynomials} \cite{Cpols,MMcpols}.
For the figure eight knot $4_1$ they are just trivial:
\be
F^{4_1}_s = 1
\ee
and
\be
a_s^{4_1}=1
\ee
thus the two branches of Kashaev limit are entirely related to the ambiguity
in treating the $\prod_j \{\ft q^{j+1}\}$ factor in (\ref{DE41}) --
as just the power of $\{\ft\}$ or as the full-fledged quasiclassical exponential.
To the contrary, the $\Cp$-polynomial for the trefoil is a little more interesting,
see (24) in \cite{MMcpols}, but there is no ambiguity in Kashaev limit.
Ambiguities increase drastically for other knots,
where non-trivial coefficients $a_s(q)$ can have different integral realizations
in Kashaev limit.
The way they are controlled and classified by $\Cp$-equations is still obscure.
An equally interesting question is how to interpret the figure-eight knot $4_1$
as a kind of a universal converter between the $\Ap$- and $\Cp$-polynomials.

\section{Conclusion}

This paper is devoted to quantum $\Ap$-polynomials -- one of the most obscure
parts of the knot theory.
It concerns the dependence of Wilson averages on representation,
and the claim is that this dependence is restricted by some kind of a finite
difference equation(s) -- just a single one if only symmetric
representation are considered.
Technically increase of representation can be described by cabling --
addition of an extra line along a knot, twisting around it in different ways.
Then this extra line can be eliminated by an application of skein relations --
thus expressing the average with increased representation through original one.
While strategically clear, this procedure is technically involved,
and its outcome is still difficult to anticipate and describe.
A more invariant statement is that a variety of links around the knot
produces a set of knot polynomials which are not all linearly independent --
and then the resultant of this system produces a relation, which is the
$\Ap$-polynomial.
Moreover, it can be convenient to further substitute links by something else,
like action of chord operators -- what is today best technical option for
the study of generic $\Ap$-polynomials \cite{GMqap}.

$\Ap$-polynomials are drastically simplified in the limit of large representations,
especially in the double scaling Kashaev limit, when simultaneously $q\longrightarrow 1$.
This is typical quasiclassics, which in application to knots we call Kashaev limit \cite{Kashaev}.
Despite this not necessary,
it is usually described as summation of specific series,
known as differential or cyclotomies expansion \cite{DE}.
As usual in quasiclassics, the sum is actually a multivalent quantity,
and suffers from Stokes ambiguities.
In the case of knots additional phenomenon is that there is a limit when the
classical action vanishes and the answer reduces to inverse determinant.
Intriguingly, it is then made from inverse Alexander(s),  as first observed by
P.Melvin and H.Morton for Jones in symmetric representations
(and generalized in \cite{KLM} to HOMFLY/Kauffman in  arbitrary ones).
In \cite{GMkl} we noted that this implies the peculiar structure of
the full (non-homogeneous) $\Ap$-polynomial equation,
which in Kashaev limit explicitly contains Alexander.
Moreover, even the homogeneous part, and even its further simplification --
the {\it classical} $A$-polynomial at $L=1$ -- is just proportional to Alexander:
$A^\CK(1,M)\sim \Al^\CK(M)$.
Since classical $A$-polynomials are well studied and known for quite many knots,
this prediction is easy to test -- and it appears true, but only partly,
probably because conventional answers for homogeneous $A$-polynomials are oversimplified
and do not correspond to {\it polynomial} non-homogeneous free terms.
This is, however, a far more difficult thing to check -- because of the lack of knowledge
about full-fledged non-homogeneous {\it quantum} $\Ap$-polynomials.

We hope that discussion and examples of the present paper will add new stimulus
for the study of this interesting subject.
As we demonstrate, there are plenty of expectations, which can be tested in every
particular example, what makes the study of these examples not pure technical
and purposeless, but somewhat conceptual.
This can attract new researchers to the field.

\section*{Acknowledgements}

I am indebted to D.Galakhov and E.Lanina for numerous explanations and comments.
This work is supported by the grant RSF 24-12-00178.


\begin{thebibliography}{12}

\bibitem{GMkl} D.Galakhov and A.Morozov, arXiv:2605.31588.


\bibitem{CS} S.-S. Chern and J.Simons, Ann.Math. 99 )1074) 48-69 \\
E.Witten, Comm.Math.Phys. 121 (1989) 351-399

\bibitem{UFN3} A.Morozov, UFN 162 \#8 (1992) 84; UFN 37 (1994) 1-55, hep-th/9303139;
 hep-th/9502091; hep-th/0502010; arXiv:2212.02632; \\
 A.Mironov, Int.J.Mod.Phys. A9 (1994) 4355; Phys.Part.Nucl. 33 (2002) 537, hep-th/9409190; Electron.
Res. Announ. AMS 9 (1996) 219-238, hep-th/9409190


\bibitem{qAp} S.Garoufalidis, Geom.Topol.Monogfr. 7 (2004) 291-309, math/0306230 

\bibitem{MMeqknpols}
A.Mironov and A.Morozov, AIP Conf.Proc. 1483 (2012) 189-211, arXiv:1208.2282

\bibitem{GMqap} D.Galakhov and A.Morozov,    
Phys.Lett.B 860 (2025) 139139, arXiv:2408.08181;
Eur.Phys.J.C 85 (2025) 8, 915, arXiv:2505.20260;  
arXiv:2605.22560


\bibitem{Ap}
D Cooper, D Long,   J. Knot Theory Ramifications 5 (1996) 609–628; \\
D W Boyd,   Experiment. Math. 7 (1998) 37–82; \\
M Culler, A table of A–polynomials Available at http://www.math.uic.edu/
~culler/Apolynomials; \\
H Murakami, J Murakami,   Acta Math. 186 (2001) 85–10; \\
R Gelca, Proc. Amer. Math. Soc. 130 (2002) 1235–1241; \\
K Habiro,   Geom. Topol. Monogr. 4 (2002) 55–68; \\
J.S.Purcell, {\it Hyperbolic Knot Theory}, arXiv:2002.12652; \\
J.E.Andersen and R.Kashaev, arXiv:1109.6295; \\
M. Aganagic, T. Ekholm, L. Ng, C. Vafa, Adv.Theor.Math.Phys. 18 (2014) 827-956, arXiv:1304.5778; \\
S. Garoufalidis, M. Goerner, and C. Zickert,
Algebraic \& Geometric Topology, 15 (2015) 371–397;\\
K. Hikami and R. Inoue,
Journal of Knot Theory and Its Ramifications, 23 (2014) 1450006;
Algebraic \& Geometric Topology, 15 (2015) 2175–2194; \\
S.Garoufalidis, D.P.Thurston and C. Zickert. 
Duke Mathematical Journal, 164 (2015) 2099–2160 

\bibitem{Jones} V.F.R.Jones,  Ann. of Math. 126 (1987) 335–38;
Discrete Math 294 (2005)   275–277 \\
L.H.Kauffman, Topology 26.3 (1987)  395–407

\bibitem{HOMFLY-PT} P. Freyd, D. Yetter, J. Hoste, W.B.R. Lickorish, K. Millett, and A. Ocneanu,
Bulletin (new series) of the American mathematical society 12.2 (1985), pp. 239–246;
J.H. Przytycki and K.P. Traczyk, Kobe J. Math.  4 (1987)  115–139, arXiv: 1610.06679 [math.GT].

\bibitem{Kauffman}  L.Kauffman, 
 Transactions of the AMS, 318 (1990) 417–471

\bibitem{Cpols}   S. Garoufalidis and X. Sun, math/0504305

\bibitem{MMcpols}
A.Mironov and A.Morozov,     JHEP 02 (2021) 142, arXiv:2009.11641

\bibitem{DE} N.M.Dunfield, S.Gukov and J.Rasmussen, Experimental Math. 15 (2006) 129-159, math/0505662 \\
E.Gorsky, S.Gukov and M.Stosic, arXiv:1304.3481 \\
A.Mironov, A.Morozov and An.Morozov, AIP Conf. Proc. 1562 (2013) 123, arXiv:1306.3197 \\
S.Arthamonov, A.Mironov and A.Morozov, Theor.Math.Phys. 179 (2014) 509-542, arXiv:1306.5682  


\bibitem{KhR}   M. Khovanov, Duke Math.J. 101 (2000) no.3, 359426, math/9908171; Experimental Math. 12 (2003) no.3,
365374, math/0201306; J.Knot theory and its Ramifications 14 (2005) no.1, 111-130, math/0302060; Al-
gebr. Geom. Topol. 4 (2004) 1045-1081, math/0304375; Int.J.Math. 18 (2007) no.8, 869885, math/0510265;
math/0605339; arXiv:1008.5084 \\
D. Bar-Natan, Algebraic and Geometric Topology 2 (2002) 337-370, math/0201043; Geom.Topol. 9 (2005)
1443-1499, math/0410495; J.Knot Theory Ramifications 16 (2007) no.3, 243255, math/0606318 \\
M. Khovanov and L. Rozansky, Fund. Math. 199 (2008), no. 1, 191, math/0401268; Geom.Topol. 12 (2008),
no. 3, 13871425, math/0505056; math/0701333 \\
N. Carqueville and D. Murfet, arXiv:1108.1081 \\
V. Dolotin and A. Morozov,
JHEP 1301 (2013) 065,arXiv:1208.4994;
J. Phys. 411 012013, arXiv:1209.5109;
Nucl.Phys. B878 (2014) 12-81, arXiv:1308.5759 \\
S. Nawata and A. Oblomkov, arXiv:1510.01795; \\
D.Galakhov, E.Lanina and A.Morozov, arXiv:2605.01584

\bibitem{Kashaev} R.Kashaev,  Lett.Math.Phys. 39 (1997) 269–275, q-alg/9601025;
Mod.Phys.Lett. A 39 (1997) 269–275


\bibitem{MM}
P.Melvin and H.Morton,  Commun.Math.Phys. 169 (1995) 501-520;\\
L. Rozansky,   Commun.Math.Phys. 175 (1996) 275–296, hep-th/9401061;\\
D. Bar-Natan and S. Garoufalidis Invent. Math. 125 (1996) 103–133; \\
L.Rozansky,  Commun.Math.Phys. 183  (1997) 291–306, arXiv:9601009; \\
S.Garoufalidis and C.Wheeler, arXiv:2603.01619

\bibitem{KLM} D.Korzun, E.Lanina and A.Morozov, to appear

\bibitem{Burau} W.Burau, 
Abh. Math. Sem. Univ. Hamburg. 11 (1936) 179–186 


\bibitem{FK} L.Faddeev and R.Kashaev, Mod.Phys.Lett. A9 (1994) 427-434

\bibitem{univR}  V.G. Drinfeld, 
Amer. Math. Soc., Providence, RI (1987)  198–820

\bibitem{Hikami} K.Hikami, Int.J.Mod.Phys. A 16 (2001) 3309-3333, math-ph/0105039;
J.Geom.Phys. 57 (2007) 1895-1940, math/0604094; \\
D.Galakhov, A.Mironov and A.Morozov, JETP, 120 (2015) 623-663, arXIv:1410.8482

 
\bibitem{RT} 
N. Reshetikhin and V. Turaev. 
Comm.Math.Phys. 127  (1990)  1–26;\\
A. Mironov, A. Morozov, and And. Morozov,  JHEP 2012.3 (2012)  1–34, arXiv: 1112.2654

\bibitem{BiMo}  L. Bishler and A. Morozov,  Phys.Lett. B808 (2020) 135639, arXiv:2006.01190 

\bibitem{Konodef}
Ya.Kononov and A.Morozov,     JETP Lett. 101 (2015) 831-834,
arXiv:504.07146

\bibitem{GK} S.Garoufalidis and Ch.Koutschan, Exp.Math. 21 (2012) 241-251, arXiv:1101.2844;
for final data see    https://www3.risc.jku.at/people/ckoutsch/pretzel/

\bibitem{evo} A,Mironov, A.Morozov and An.Morozov, AIP Conf. Proc. 1562 (2013) 123,   arXiv:1306.3197
  
\bibitem{Cooper} D.Cooper, M.Culler, H.Gillet, D.D.Long  and P.B.Shalen, Invent. math. 118 (1994) 47-84 



\bibitem{katlas} http://katlas.org


\end{thebibliography}
\end{document}